\documentclass[12pt,preprint]{aastex}

\begin{document}

\title{A direct method for measuring heat conductivity in intracluster medium} 

\author{Makoto Hattori\email{hattori@astr.tohoku.ac.jp} and Nobuhiro Okabe\email{okabe@astr.tohoku.ac.jp}}
\affil{Astronomical Institute, Graduate School of Science, Tohoku University,\\
Sendai 980-8578, Japan.}


\begin{abstract}

The inverse Compton scattering of the cosmic microwave background (CMB) radiation  
with electrons in the intracluster medium which has  a temperature gradient,
was examined by the third-order perturbation theory of the Compton scattering. 
A new type of the spectrum distortion of  the CMB was found and named 
as gradient T Sunyaev-Zel'dovich effect (gradT SZE). 
The spectrum has an universal shape. 
The spectrum crosses over zero at 326GHz. 
The sign of the spectrum depends on the relative direction of the
line-of-sight to the direction of the temperature gradient. 
This unique spectrum shape can be used to detect the gradT SZE signal 
by broad-band or multi-frequency observations of the SZE. 
The amplitude of the spectrum distortion does not depend 
on the electron density and is proportional to  
the heat conductivity.  
Therefore, the gradT SZE provides an unique opportunity 
to measure thermally nonequilibrium 
electron momentum distribution function when the ICM has 
a temperature gradient and  the 
heat conductivity in the ICM. 
However, the expected amplitude of the signal is very small. 
The modifications to the thermal SZE spectrum due to variety of 
known effects, such as relativistic correction etc., 
can become problematic when using multi-frequency separation techniques to 
detect the gradT SZE signal. 

\end{abstract}



\section{Introduction}

The recent dramatic progresses of the X-ray observations  
have been  unveiling that the temperature distribution of the 
intracluster medium (hereafter ICM) is far from isothermal. 
It has turned out that the 
heat conductivity is one of the key parameters 
controlling the evolution of the ICM(Ikebe et al. 1999; Markevitch et al. 2000; Fabian et al. 2001;
Vikhlinin et al. 2001; Markevitch et al. 2003; Zakamska \& Narayan 2003).    
The global temperature gradient have been found  in some cooling core clusters\cite{Kaa04}.
Zakamska \& Narayan (2003) performed hydrostatic model fitting of the 
observed temperature and density profile of the ICM.
They assumed the energy balance between the radiative cooling and 
the conductive heating. 
They showed that the heat conductivities should be 
one-third of the Spitzer value in majority of the clusters.
Shock waves\cite{Mark02,Fujita04}  and cold fronts(Markevitch et al. 2000; Vikhlinin et al. 2001)
found in merging clusters are showing the temperature jumps 
across the discontinuities.
The hot bubbles caused by the interaction with  
the activity of the central galaxies
have been  found in many clusters\cite{McN00,Fab00,Blan01}.
The heat conductivities across the cold fronts and 
the surface of the hot bubbles have to be 
reduced in many order of magnitudes to maintain the structures\cite{Ett00}. 
The large scale fluctuation of the temperature distributions have been 
found in several clusters(Watanabe et al, 1999; Shibata et al. 2001; Markevitch et al. 2003). 
To maintain the observed temperature fluctuations at least for a dynamical time
scale of clusters, the order of magnitude reduction of the heat conductivities 
are required\cite{Mark03}.     
All the above mentioned facts show that 
a direct method for measuring the heat conductivities 
may provide crucial informations to study the evolution of the ICM.

The recent theoretical progresses have been unveiling that 
a microscopic instability of the plasma which has a temperature gradient\cite{RL78}, 
could play important roles on 
the suppression of the heat conductivities\cite{lev92,hat00,OH03} 
and the origin  of cluster magnetic fields(Okabe \& Hattori 2003; Okabe \& Hattori 2004).
The heat conduction is the electron thermal energy flow from hot to cold regions. 
Since it is the third order 
moment of the electron momentum distribution function,  
the electron momentum distribution function must have deviation from the 
Maxwell-Boltzmann distribution which is the even function of the momentum. 
Therefore, the electron momentum distribution function of the plasma with a temperature gradient
is described by the sum of the Maxwell-Boltzmann distribution, 
$g_{\rm m}({\bf q})=(2\pi)^3 n_e 
(2\pi m_ek_BT_e)^{-3/2}$  ${\rm exp}\left(-{q^2\over 2m_e k_BT_e}\right)$, and 
the non Maxwellian part, $\Delta g({\bf q})$, where ${\bf q}$ is a momentum of 
electron, and $n_e$, $T_e$ and $m_e$ are electron number density, temperature 
and  mass, respectively. 
The heat current density, $\kappa {\bf\nabla}T_e$, 
is described by the $\Delta g({\bf q})$ as 
\begin{eqnarray*}
\int {d^3{\bf q}\over (2\pi)^3}{1\over 2m_e}q^2 {{\bf q}\over m_e}\Delta g({\bf q})&=&
-f_{\kappa}\kappa_{\rm Sp} {\bf\nabla}T_e,
\end{eqnarray*} 
where $f_{\kappa}$ is the heat conductivity 
in the Spitzer value unit and 
$\kappa_{\rm Sp}$ is the Spitzer heat conductivity. 
The Spitzer heat conductivity is given by 
$\kappa_{\rm Sp}=1.31\lambda_e n_e k_B$  $\left({k_B T_e\over m_e}\right)^{1/2}$ where
$\lambda_e$ is the electron Coulomb mean free path\cite{Sp56,sar88}.
The deviation of the electron momentum distribution from the thermal equilibrium 
distribution\cite{OH03} in the  plasma with a temperature gradient is described by 
\begin{eqnarray}
\Delta g({\bf q})&=&f_{\kappa} \kappa_{\rm Sp}{m\over k_B n_e}
{{\bf q}\cdot {\bf \nabla}T_e\over 
q_{th}^2 T_e}\left(2- {4q^2\over 5 q_{th}^2}\right) g_{\rm m},
\end{eqnarray}
where  $q_{th}=\sqrt{2 m_e k_B T_e}$ is the electron thermal momentum. 
The plasma instability is derived by this local deviation of the electron 
momentum distribution from the thermal equilibrium distribution.
It has been shown that the instability may provide an origin of the 
magnetic fields in cluster of galaxies\cite{OH03}.
Therefore, the measurement of  the functional form of the non-equilibrium 
momentum distribution function is crucial for 
confirming the above mentioned theoretical models on the suppression of the
heat conductivities and generation of the magnetic fields.

In this paper, a spectrum distortion of the cosmic microwave background
(hereafter CMB) due to the inverse  Compton scattering with electrons 
which support heat conduction, is examined 
by solving the Boltzmann equation for Compton scattering to third-order 
in the electron momentum. 
The result provides a direct method for measuring the heat conductivities and 
the functional form of the electron momentum distribution functions in the 
plasma with a temperature gradient.

\section{The Boltzmann equation for third-order Compton scattering}

The spectrum distortion of the cosmic microwave background (CMB) due to 
the Compton scattering from  electrons with  momentum distribution function 
of $g(\bf{q})$ is given by 
the Boltzmann equation\cite{Hu94} as
\begin{eqnarray}
{D f({\bf x},{\bf p})\over cDt}&=&{1\over 2E({\bf p})}\int D{\bf q}D{\bf q'}D{\bf p'}
(2\pi)^4\delta^{(4)}(\vec{p}+\vec{q}-\vec{p'}-\vec{q'})|M|^2\nonumber\\
&&\times \{g({\bf x},{\bf q'})f({\bf x},{\bf p'})[1+f({\bf x},{\bf p})] -
g({\bf x},{\bf q})f({\bf x},{\bf p})[1+f({\bf x},{\bf p'})]\},
\end{eqnarray}
where $f({\bf x},{\bf p})$ is the photon occupation number with a three-momentum ${\bf p}$, 
${\bf p}$ and ${\bf p'}$ are the three-momentum for photons of four-momentum $\vec{p}$ and $\vec{p'}$,  ${\bf q}$ and ${\bf q'}$ are
the three-momentum for electrons of four-momentum $\vec{q}$ and $\vec{q'}$, $E({\bf p})$ is the energy of the photon with a momentum 
${\bf p}$, $|M|^2$ is the Lorentz-invariant 
matrix element for
Compton scattering\cite{Hu94},
and 
$$D{\bf q}={d^3{\bf q}\over (2\pi)^3 2 E({\bf q})}$$
is the Lorentz-invariant phase space volume element of electrons with
an energy of $E({\bf q})$  
and  a momentum ${\bf q}$. 
The left hand side of the equation is the total derivative of the 
photon distribution function in the phase space written as,
\begin{eqnarray*}
{D f\over D t}&=&{\partial f\over \partial t}+{\partial f\over \partial x^i}{d x^i\over d t}+{\partial f\over \partial n_i}{d n_i\over d t} +{\partial f\over \partial E(p)}{d E(p)\over d t},
\end{eqnarray*}
where $n_i$ is  the $i$-th component of the direction cosines, ${\bf n}$,   for a photon of four-momentum $\vec{p}$. 
The first two terms in the right hand side is a derivative along the photon path. The third term represents an effect of gravitational deflect of the photon path due to the cluster gravitational lensing effect. 
Since the deflection angle due to the cluster gravitational lensing effect is at most $30''$, the change of the path length due to the lensing effect is at most 0.01pc and negligibly small compared with the cluster size of $\sim$1Mpc. 
The last term represents the cosmological redshift and  the energy change of the photons due to the gravitational scattering 
by the tangential peculiar motion of the cluster\cite{Birk84,aso02}.
The gravitational scattering  causes the CMB temperature fluctuation of  $\delta T/T\sim 10^{-7}$\cite{aso02}. The effect is order of magnitude smaller than the Compton scattering effect concerned in this paper as shown in the next section.   
Therefore, the terms representing the gravitational lensing effect and the gravitational scattering effect are neglected in the course of the following calculations.  
Further, the cosmological  redshift effect is neglected for simplicity 
since the effect is trivial\cite{Hu94}.

The right hand side of the Boltzmann equation is the collision term.
The first and the second terms are the contributions from scattering into and out of the momentum state ${\bf p}$ including the stimulated emission effects, respectively.  
In the rest frame of the incident electron, the matrix element for
Compton scattering summed over polarization is given by,
\begin{eqnarray*}
|M|^2&=&2(4\pi)^2 e^2\left[{\tilde{p'}\over \tilde{p}}+{\tilde{p}\over\tilde{p'}}-1+(\tilde{\bf n}\cdot\tilde{\bf n'})^2\right],
\end{eqnarray*}
where $e$ is the absolute value of the charge of the electron.
The variables with $\sim$ denotes those in the electron rest frame, 
otherwise those in the laboratory frame.

As an example, the case 
when the electron velocity distribution function is  
the Maxwell-Boltzmann and the thermal energy contained in 
electron system is much less than the electron rest mass
energy, is considered. 
By expanding the collision term of the Boltzmann equation for 
isotropically distributed incident photons 
up to the second-order in electron momentum, the Kompaneets equation is
deduced\cite{Hu94,Komp57}.  
The spectrum distortion of the CMB due to the Compton scattering 
from thermal electrons contained in the clusters of galaxies 
are obtained by solving the Kompaneets equation under the conditions
that the probability of the 
multiple scattering of a photon is negligibly small and the electron 
temperature is much higher than the temperature of the CMB.  
This effect is known as the thermal Sunyaev-Zel'dovich effect (hereafter thermal SZE) 
\cite{SZ72}.

The spectrum distortion due to the plasma with a temperature gradient is deduced by the 
following way. 
In the plasma with a temperature gradient, the electron momentum distribution function 
is described as $g({\bf q})=g_{\rm m}({\bf q})+\Delta g({\bf q})$.  
Since the multiple scattering of a photon is negligible in the clusters of 
galaxies, the effects of the spectrum distortion of the CMB due to the Compton scattering
from electrons belonging to $g_{\rm m}(\bf{q})$ and electrons belonging to
$\Delta g({\bf q})$ 
are decoupled.  
Therefore, each effects can be evaluated separately and the total spectrum distortion 
is given by the superposition of each results. 
Hereafter, the SZE due to electrons belonging to $\Delta g({\bf q})$ is referred to 
gradT SZE. 
Since the first non-zero moment of 
the  $\Delta g({\bf q})$ appears from the third-order moment, 
to evaluate the effect of the  $\Delta g({\bf q})$ on the CMB spectrum distortion
the third order expansion in electron momentum of the collision term of
the Boltzmann equation is required. 
The technical details are summarized in Appendix. 
A final result of third-order expansion is reduced to 
\begin{eqnarray}
{D f({\bf x},{\bf p})\over cDt}&=&3f_{\kappa}\kappa_{\rm Sp}{\sigma_{\rm T}\over mc^3} 
{\bf n}\cdot{\bf \nabla}T_e\nonumber\\ 
&\times&\left[ xf'({\bf x},{\bf p})
+{94\over 75}x^2f''({\bf x},{\bf p})
+{14\over 75}x^3 f'''({\bf x},{\bf p})\right],
\end{eqnarray}
where $x={h\nu\over k_B T_{\gamma}}$, $h$ is the Planck constant, $c$ is a speed of light,
$\nu$ is the frequency 
of the CMB photon, $T_{\gamma}$ is the temperature of the CMB, $\sigma_{\rm T}$ is the
Thomson cross section and ${\bf n}\equiv {{\bf p}\over p}$ is the direction cosine 
of the observed photon, respectively. 
Although the resultant photon distribution is anisotropic, 
the isotropic photon distribution for incident photons is assumed in the course of calculation
since the multiple scattering is negligible and the incident photons are 
CMB photons.  
The quantities $k_BT_e/mc^2$ and $E({\bf p})/mc^2$ are treated as the order of $(v/c)^2$ 
variables.

\section{The spectrum of the gradient-T SZE}

The spectrum of the gradT SZE 
is obtained  
by inserting the Planck function in the right hand side of the above obtained 
third-order Boltzmann equation 
and integrating over line of sight. 
The frequency dependent amplitude of the distortion expressed in temperature is 
obtained as 
\begin{eqnarray}
\Delta T^{\nabla T}_{\nu}&=&{12\sigma_{\rm T}\over m c^3}
T_{\gamma}\int d\ell f_{\kappa}\kappa_{\rm Sp}{\bf n}
\cdot {\bf \nabla}T_e{1\over x^2}\Delta^{\nabla T}(x),\\
\Delta^{\nabla T}(x)&\equiv&-{x^4 {\rm e}^x\over ({\rm e}^x-1)^2} 
\left[ 1
-{94\over 75} x {\rm coth}{x\over 2}+{14\over 75} x^2 \left({\rm coth^2}{x\over 2} 
+{1\over 2 {\rm sinh^2}{x\over 2}}\right)\right],
\end{eqnarray}
where $d\ell$ is the line of sight integral. 
Figure 1 compares the spectrum shape of the gradT SZE with the 
thermal and kinetic SZE\cite{Hu94,Birk99}.
Figure 2 shows the frequency dependence of the kernel function $\Delta^{\nabla T}(x)$
to see diagnostics of the spectrum shape of the 
gradT SZE compared with the thermal and kinetic SZE\cite{Hu94,Birk99}. 
The spectrum shows distinct characteristics compared with others.
Therefore, the  functional form of 
the non-Maxwellian part can be measured by 
this spectrum shape of the gradT SZE.  
It has a finite value at the cross over frequency of the thermal SZE, 
that is 218GHz. 
The location of maxima, cross over frequency, and minima are 
independent of $T_e$. 
They are at $x_{\rm max}=3.44\;(\nu=195{\rm GHz})$, $x_{\rm zero}=5.75;
(\nu=326{\rm GHz})$, and $x_{\rm min}=8.36;(\nu=437{\rm GHz})$, respectively.
Since the amplitude of the spectral distortion is proportional to 
the inner product of the ${\bf n}$ with ${\bf \nabla}T_e$,  
the sign of the spectrum depends on the relative direction of the 
line-of-sight to the temperature gradient. 
When the hotter region locates closer to the observer, 
the amplitude of the distorted spectrum is larger than the CMB 
in the Rayleigh-Jeans part. 
A spherically symmetric system does not  
show the spectrum distortion due to the gradT SZE 
since both of increment and decrement with the same amount appear
in a single line-of-sight and these are cancelled out. 
When the line of sight is exactly perpendicular to the temperature gradient, 
there is no gradT SZE effect. 

The amplitude of the spectrum distortion due to the gradT SZE 
at the peak frequency is evaluated by the following way. 
The line of sight integral in the right hand side of the equation 
is reduced to $T_e^{7/2}$.  It is independent from the 
electron density because the heat conductivity does not 
depend on the electron density explicitly unless it is saturated\cite{CM77}.
As decreasing the electron density the Coulomb mean free path becomes 
longer. In the region where the electron density is lower than $10^{-5}{\rm cm^{-3}}$ with $k_B T_e=10$keV, the mean free path is longer than 2Mpc\cite{sar88}. 
For the typical rich cluster, the electron density gets lower than $10^{-5}{\rm cm^{-3}}$ in the region of  2Mpc apart from the cluster center. 
Therefore, in such a  cluster out skirt the Coulomb mean free path is 
longer than temperature gradient scales 
and the heat conduction is 
saturated.  
As modeled by the Cowie and Mackee (1977), the saturated heat conductivity could be evaluated by the collisionless diffusion of the hot electrons 
with the Maxwellian velocity distribution. 
It is reasonable to assume that 
the electron velocity distribution  is the Maxwellian
in the cluster out skirt region where the heat conduction is saturated
since the deviation of the velocity 
distribution function from the thermal equilibrium form is hard to be
established without any collision between particles.  
Therefore, we assume that  $\Delta g=0$ in the cluster out skirt region 
where the heat conduction is saturated.  
For simplicity,  we assume that 
$\Delta g$ disappears suddenly at the points where the electron
Coulomb mean free path becomes as long as the temperature gradient scale.  
The temperature of these transition 
points for the closer side and far side to the 
observer along the line-of-sight are denoted by $T_{e2}$ and $T_{e1}$, 
respectively.
Then, the amplitude of the gradT SZE is obtained as 
\begin{eqnarray}
\Delta T^{\nabla T}_{\nu}&=&0.8\left({f_{\kappa}\over 1.0}\right)
\left[{(k_BT_{e2})^{3.5}- 
(k_BT_{e1})^{3.5}\over (10{\rm keV})^{3.5}-(5{\rm keV})^{3.5}}\right]
\mu {\rm K}.
\end{eqnarray}
The result shows that the amplitude of the gradT SZE is proportional to 
the heat conductivity.
To extract the heat conductivity from the observed amplitude of 
the gradT SZE, the electron temperature
in the cluster out skirt at the both side of the line-of-sight 
have to be measured. 
Since it is practically difficult, some assumptions must be made 
to extract the heat conductivity from the observed temperature map. 
The expected amplitude is two order of magnitude less than the thermal SZE
and an order of magnitude less than the kinetic SZE 
but order of magnitude larger 
than the gravitational scattering effect. 
To distinguish the gradT SZE signal from other SZE signals, 
measurement of the spectrum by broad band or multi-band observations 
are required.

\section{Applications to clusters of galaxies}

The applications of the gradT SZE to three different situations 
in clusters of galaxies 
are considered in this section. 
First case is that the plasma has a smooth distribution 
of density and temperature.   
Second and third cases contain one discontinuity, such as a shock and 
a cold front along the line-of-sight, respectively.  
An idealized line-of-sight passing through clusters of galaxies 
is illustrated in Figure 3.
The region denoted by $D$ is a discontinuity.
The regions denoted by 1 and 2 are smooth regions. 
The outer edges of these regions are defined by the place
where the heat conduction 
starts to be saturated (Cowie \& McKee 1977; Sarazin 1988).  
For simplicity, the reduction rate of the heat conduction in these 
smooth regions, $f_{\kappa}$, is assumed to be constant through out
the regions and 
suddenly becomes zero at each edges.
The electron temperature at the edge of the discontinuity and at the outer 
edge of each smooth regions are denoted by suffixes $D$ and $o$, 
respectively. 
Since the scale lengths of the discontinuities, $L_D$, are 
comparable to or shorter than the Coulomb mean free path,  
the reduction rate of the heat conduction in the discontinuities,
$f_{\kappa}'$, may be different from 
that in the smooth region and could be much less than 1.  

The general expression for the 
amplitude of the gradT SZE for the observer illustrated in Figure 3 
is obtained by performing the line-of-sight integral as 
\begin{eqnarray*}
\Delta T_{\nu}^{\nabla T}&\propto& f_{\kappa} \left[T_e^o(2)^{7/2}-T_e^D(2)^{7/2}\right] \\
&&+f_{\kappa}'\left[T_e^D(2)^{7/2}-T_e^D(1)^{7/2}\right]
-f_{\kappa} \left[T_e^o(1)^{7/2}-T_e^D(1)^{7/2}\right].
\end{eqnarray*}

{\bf No discontinuity along a line of sight}

When there is no discontinuity along a line of sight, the reduction rate 
of the heat conductivity is constant through out the line of sight. 
By setting $f_{\kappa}'=f_{\kappa}$, the amplitude of the gradT SZE 
at $\nu=195$GHz is reduced to 
\begin{eqnarray*}
\Delta T_{\nu}^{\nabla T}&\sim & 0.02 \left({f_{\kappa}\over 0.3}\right) 
\left[{T_e^o(2)^{7/2}-T_e^o(1)^{7/2}\over (5{\rm keV})^{7/2}-(3{\rm keV})^{7/2}}\right]\mu {\rm K}.
\end{eqnarray*}
The amplitude depends on the temperature at both outer edges of
the smooth region  
and the heat conductivity in the smooth 
region. 
The amplitude is insensitive to any local variation of the temperature
along line-of-sight. 
When the temperatures at the both outer edges are close to each other, 
the amplitude of the gradT SZE becomes zero. 

{\bf Across a shock front}

Although the electron distribution function in the cluster shock waves are 
quit uncertain, it is likely that the shock is collisionless and the thickness 
of the shock front is much smaller than the electron mean free path. 
Therefore, it is expected that 
the heat conductivity in the shock waves is much less than 
that in the smooth region, e.g. $f_{\kappa}'\ll f_{\kappa}$. 
The amplitude of the gradT SZE across a shock front at $\nu=195$GHz 
is reduced to
\begin{eqnarray*}
\Delta T_{\nu}^{\nabla T}&\sim & -12 \left({f_{\kappa}\over 0.3}\right) 
\left[{T_e^D(2)^{7/2}-T_e^D(1)^{7/2}\over (30{\rm keV})^{7/2}-(5{\rm keV})^{7/2}}\right]\mu {\rm K},
\end{eqnarray*}
where the effect comes from the 
difference of the  electron temperatures 
at the both sides of the outer edges in the smooth regions is neglected. 
The adopted temperature of 30keV as the post shock gas temperature 
is somewhat larger than the average 
X-ray temperature of the cluster hot gas, that is 2-10keV. 
However, there are several observational and theoretical 
evidences which indicate that the temperature of 
the shock heated gas due to a cluster-cluster major merger could be 
as hot as 30keV. 
A cluster 1E0657-56 is an on-going merging cluster in which 
a contact discontinuity and a shock are identified (Markevitch et al. 2002). 
The temperature map of this cluster indicates the existence of the 
high temperature region hotter than 20keV. 
The combined analysis of the 
multi wave band SZE mapping observations of the most X-ray luminous 
cluster RXJ1347-1145 (Kitayama et al. 2004) showed that the inferred 
temperature of the south-east part of the cluster is well in 
excess of 20keV  which is confirmed by the Chandra X-ray temperature
measurement (Allen et al. 2002). 
A numerical simulation of the head on collision of rich clusters (Takizawa 1999) showed  that the shock heated gas temperature 
may temporary exceed 20keV. 
Therefore, in such an ideal case, the amplitude of the gradT SZE could be 
comparable to that of the kinetic SZE.
An important point we have to emphasize is that the amplitude of the gradT SZE across the shock front  is proportional to the heat conductivity in the smooth region. 
Therefore, this observation can be used to measure the heat conductivity in 
the smooth region. 

{\bf Across a cold front}

The thickness of cold fronts are yet unresolved.
The current observed upper limits are 
comparable to the electron mean free path. 
Since the cold fronts are found in relatively large fraction of clusters,
the cold fronts should not be a short lived transient structure. 
To maintain the cold fronts, the heat conductivity should be suppressed in 
many order of magnitude. 
It is likely that the heat conductivity in
cold fronts is much less than unity, $f_{\kappa}'\ll 1$, and
much less than that in the smooth region, $f_{\kappa}'\ll f_{\kappa}$.
Therefore, the amplitude of the gradT SZE when the line of sight
acrosses one cold front, is estimated as 
\begin{eqnarray*}
\Delta T_{\nu}^{\nabla T}&\sim & -0.1 \left({f_{\kappa}\over 0.3}\right) 
\left[{T_e^D(2)^{7/2}-T_e^D(1)^{7/2}\over (8{\rm keV})^{7/2}-(4{\rm keV})^{7/2}}\right]\mu {\rm K},
\end{eqnarray*}
where the effect comes from the 
difference of the  electron temperatures 
at the both sides of the outer edges in the smooth region is neglected.
The dependence on the physical variables are the same as the case acrossing 
a shock. 
However, 
the temperature increase acrossing the cold front is 
not so significant since there is no thermalization of
the kinetic energy as in the case of the shock.  
Therefore, the expected amplitude is much  smaller than the case of acrossing
the shock.

\section{Discussion}

We have  
shown that the inverse Compton scattering of the CMB photon 
with electrons in the ICM which has a temperature gradient, 
results in a new type of the spectrum distortion of  the CMB
by the third-order perturbation theory of the Compton scattering. 
The spectrum has an universal shape.    
The cross over frequency appears at 326GHz. 
The sign of the spectrum depends on the relative direction of the
line-of-sight to the direction of the temperature gradient. 
When the hotter region locates closer to the observer,
the intensity becomes brighter than   the 
CMB spectrum in the frequency regions lower than 
the cross over frequency. 
Therefore, the effect is distinguishable from other SZE effects 
by broad band or multi-bands observations of the SZE. 
The precise measurement of the spectrum shape of the gradT SZE 
provides an unique opportunity to measure the non-Maxwellian part 
of the electron velocity distribution function when the ICM has a 
temperature gradient.
The amplitude of the spectrum distortion is proportional to the heat 
conductivity.
It provides a possibility for a direct measurement of the  
heat conductivity. 
The expected amplitude of the gradT SZE signal is more than 
two order of magnitude smaller than the thermal SZE.
The biggest signal for gradT SZE  occurs when the 
line-of-sight across a shock front. 
The expected signal for this case 
is as large as $\sim 30\mu$K when $f_{\kappa}=1$ at maximum.  
This is  $\sim 10\%$ of the thermal SZE.
However, there are variety of the effects 
which distort the thermal SZE spectra, such as 
the relativistic correction (Birkinshaw 1999), the existence of the non-thermal electrons (Birkinshaw 1999), the hydrodynamical effect of the 
cluster merger (Koch 2004)
and the turbulent motion giving rised by merger shock (Nagai, Kravtsov \& Kosowsky 2003). 
These can become problematic when using multi-frequency separation techniques.

The gradT SZE does not cause any polarization since the quadrapole moment 
of the CMB intensity distribution averaged over all the electron momentum
is zero.  
The linear polarization of the Compton scattered photons is generated by the 
systematic transverse motion of the electron system since
it leads a finite quadrapole moment of the CMB intensity distribution 
in the frame moving with the same velocity as the transverse motion 
of the electron system\cite{SZ80}. 
As explained in the Section 1 and appendix,
the first and second moments of the electron momentum 
on $\Delta g({\bf q})$ are zero\cite{OH03}.
Therefore, the $\Delta g({\bf q})$ does not 
generate a quadrapole moment of the CMB intensity distribution.

The third order Doppler effect could provide  
the physical explanation of the spectrum distortion in the gradT SZE
although it is not satisfactory enough.
The scattering can be treated as almost elastic in the electron 
rest frame, i.e., $\tilde{p}'=\tilde{p}$.   
The transformation into the laboratory frame induces a 
Doppler shift and an energy transfer to the photon amount of 
$\delta p$ given as 
\begin{eqnarray*}
{\delta p\over p}&=& {1-\vec{\beta}\cdot{\bf n}\over 
1-\vec{\beta}\cdot{\bf n'}} -1
\end{eqnarray*}
Averaging over the incoming photon direction $\vec{n}$ which is isotropically 
distributed, we obtain 
\begin{eqnarray*}
\left<{\delta p\over p}\right>&\sim & \vec{\beta}\cdot{\bf n'}+(\vec{\beta}\cdot{\bf n'})^2+ (\vec{\beta}\cdot{\bf n'})^3.
\end{eqnarray*}
The averaging over electron velocity with velocity distribution function of 
$\Delta g$ results in 
\begin{eqnarray*}
\left<{\delta p\over p}\right>&\sim &-{6\over 5 n_e}f_{\kappa}\kappa_{SP} {{\bf n}\cdot\vec{\nabla} T_e \over m_e c^3}\sim \pm 0.14 \left({\lambda_e\over L}\right) 
{v_{th}\over c} {4k_B T\over m_e c^2},
\end{eqnarray*}
where $\lambda_e$ is the Coulomb mean free path of electron, $L$ is the temperature
gradient scale length and $v_{th}=\sqrt{2k_B T/m_e}$ is the electron thermal 
velocity.     
This is an 
averaged fractional energy shift of the  photon  by a single scattering. 
When the hot region locates foreground of the cold region along the 
line of sight, the minus sign in the last equality is taken.  
The scattered photons lose their energy.  
Therefore, the original Planck spectrum shifts toward low frequency region 
contrary to the case of the thermal SZE. 
This results in the increment in the Rayleigh-Jeans part and the decrement 
in the Wien part contrary to the thermal SZE.  
When the cold region locates foreground of the hot region,
the plus sign in the last equality is taken. The scattered photons 
gain the energy. 
In this case, the original Planck spectrum shifts toward high 
frequency region. The resultant spectrum distortion is the decrement
in the Rayleight-Jeans part and the increment in the Wien part
as same as the case of thermal SZE.

\section*{Acknowledgments}
The authors would like to thank Y.Fujita, T.Kitayama and N.Sugiyama for their valuable comments. 
MH would like to thank M.Hoshino for motivating him to do this work in the course of the discussion. 
The authors would also like to thank anonymous referee who provides 
constructive and objective comments which were helpful for improving the paper. 
This work is supported by a Grant-in-Aid for the 21st Century COE
Program ``Exploring New Science by Bridging Particle-Matter Hierarchy''
in Tohoku University, funded by the Ministry of Education, Science,
Sports and Culture of Japan and a Grant-in-Aid No.16204010 funded by Japan 
Society for the Promotion of Science.

\section*{Appendix}

Some technical details for performing the third order perturbation 
theory for the Compton scattering are summarized. In the following discussion, $p=E({\bf p})$ and $p'=E({\bf p'})$ are used for simplicity. 

The third order expansions of each terms contained 
in the collision integral are obtained as follows. 
The expansion to third order of the energy conservation can be treated as  an expansion in $\Delta p=E({\bf q})-E({\bf q'})$ as 
\begin{eqnarray*}
\delta(p+q-p'-q')&=&\delta(p'-p-\Delta p)\\
&=&\delta(p-p')-\Delta p {\partial\over \partial p'}\delta (p'-p)\\
&&+{1\over 2}\Delta p^2 {\partial^2\over \partial p'^2}\delta(p'-p)
-{1\over 6}\Delta p^3 {\partial^3\over \partial p'^3}\delta(p'-p).
\end{eqnarray*}
Since the electron momentum distribution function does not depends on ${\bf q}$ in the first term of the collision integral, first performing the integral by ${\bf q}$ and ${\bf q}$ is replaced with ${\bf q'}+{\bf p'}-{\bf p}$ by using the momentum 
conservation. 
Then, $\Delta p$ is written as
\begin{eqnarray*}
\Delta p&=&{1\over E({\bf q'})+p'}\left({\bf p'}\cdot{\bf (p'-p)}+{\bf q'}\cdot{\bf (p'-p)}\right).
\end{eqnarray*}
For the second term in the collision integral, 
the integral by ${\bf q'}$ is performed first.  
The energy difference $\Delta p$ is written as
\begin{eqnarray*}
\Delta p&=&{-1\over E({\bf q})+p}\left({\bf p}\cdot{\bf (p-p')}+{\bf q}\cdot{\bf (p-p')}\right).
\end{eqnarray*}
In the pre-scattering electron rest frame, 
the energy of the scattered photon is described  by the energy of the 
incident electron as 
\begin{eqnarray*}
\tilde{p}'&=&{\tilde{p}\over 1+{\tilde{p}\over m_e c^2}(1-{\bf n}\cdot{\bf n}')}. 
\end{eqnarray*}
The Doppler formula provides a relation between the photon energy in the rest frame and that in the laboratory frame as
\begin{eqnarray*}
\tilde{p}=\gamma p(1-\mbox{\boldmath$\beta$}\cdot{\bf n}),& & \tilde{p}'=\gamma p'(1-\mbox{\boldmath$\beta$ }\cdot {\bf n}'),
\end{eqnarray*}
where $\gamma$ is the Lorentz factor and $\mbox{\boldmath $\beta$}={\bf q}/mc$ is the incident 
electron velocity in the laboratory 
frame relative to the speed of light. 
The Lorentz invariant relation of $\vec{\tilde p}\cdot \vec{\tilde p'}=\vec{p}\cdot\vec{p'}$ provides a relation of the photon scattering angle 
between the rest frame and the laboratory frame as
\begin{eqnarray*}
\tilde{\bf n}\cdot\tilde{\bf n'}&=&
{-\mbox{\boldmath$\beta$}\cdot({\bf n}+{\bf n}')+(\mbox{\boldmath$\beta$}\cdot {\bf n})(\mbox{\boldmath $\beta$}\cdot {\bf n}') +
\beta^2 (1-{\bf n}\cdot{\bf n}')+{\bf n}\cdot{\bf n}'
\over 1-\mbox{\boldmath $\beta$}\cdot({\bf n}+{\bf n}')+(\mbox{\boldmath $\beta$}\cdot{\bf n})(\mbox{\boldmath $\beta$}\cdot{\bf n}')}.
\end{eqnarray*}
Since the ratio of the photon energy to the electron rest mass energy is order of $\beta^2$, the third order expansion of $|M|$ in $\beta$ is expressed as
\begin{eqnarray*} 
|M|^2&=&2(4\pi)^2 e^2 (1+\tilde{\bf n}\cdot\tilde{\bf n'})\\
&=&2(4\pi)^2e^2 [1+({\bf n}\cdot{\bf n}')^2-2\mbox{\boldmath $\beta$}\cdot({\bf n+n'}){\bf n}\cdot{\bf n'}(1-{\bf n}\cdot{\bf n'})\\
&&+(\mbox{\boldmath $\beta$}\cdot({\bf n+n'}))^2 (1-3{\bf n}\cdot{\bf n'})(1-{\bf n}\cdot{\bf n'})
+2\beta^2 {\bf n}\cdot{\bf n'}(1-{\bf n}\cdot{\bf n'})\\
&&+2(\mbox{\boldmath$\beta$}\cdot{\bf n})
(\mbox{\boldmath$\beta$}\cdot{\bf n'}){\bf n}\cdot{\bf n'}(1-{\bf n}\cdot{\bf n'})
 +2 (\mbox{\boldmath $\beta$}\cdot({\bf n+n'}))^3(1-2{\bf n}\cdot{\bf n'})(1-{\bf n}\cdot{\bf n'})\\
&&-2(\mbox{\boldmath $\beta$}\cdot({\bf n+n'}))\beta^2(1-{\bf n}\cdot{\bf n'})(1-2{\bf n}\cdot{\bf n'})\\
&&-2(\mbox{\boldmath $\beta$}\cdot({\bf n+n'}))(\mbox{\boldmath$\beta$}\cdot{\bf n})(\mbox{\boldmath$\beta$}\cdot{\bf n'})(1-3{\bf n}\cdot{\bf n'})(1-{\bf n}\cdot{\bf n'})].
\end{eqnarray*}
The third order expansion of the electron energy is 
$E({\bf q})\sim mc^2 (1+ q^2/2m^2c^2)$.

The 1st and 2nd moments of electron momentum for the $\Delta g$ are zero,
and the 3rd moments are obtained as
\begin{eqnarray*}
\int {d^3{\bf q}\over (2\pi)^3}q_iq_jq_k\Delta g&=&-{2\over 5}m^2f_{\kappa}\kappa_{\rm SP} \partial_x T_e,
\;\;\;\;{\rm for}\;\;i=x,j=k=y\;{\rm or}\;z,\\
&=&-{6\over 5}m^2f_{\kappa}\kappa_{\rm SP}\partial_x T_e,\;\;\;\;{\rm for}\;\;
i=j=k=x\\
&=&0,\;\;\;\;\;\;\;\;\;\;\;\;\;\;\;\;\;\;\;\;\;\;\;\;\;\;\;\;\;\;\;\;\;\;\;{\rm otherwise}
\end{eqnarray*} 
where the $x$ axis is taken to 
the direction of the temperature gradient. 
The non-zero terms in the third order expansion of the collision term
are only the third order moment of electron momentum and can be 
evaluated by the above results. 
The evaluations of the each terms are straight forward.

\begin{figure}[htbp]
\begin{center}
\includegraphics[angle=-90,width=1.0\linewidth]{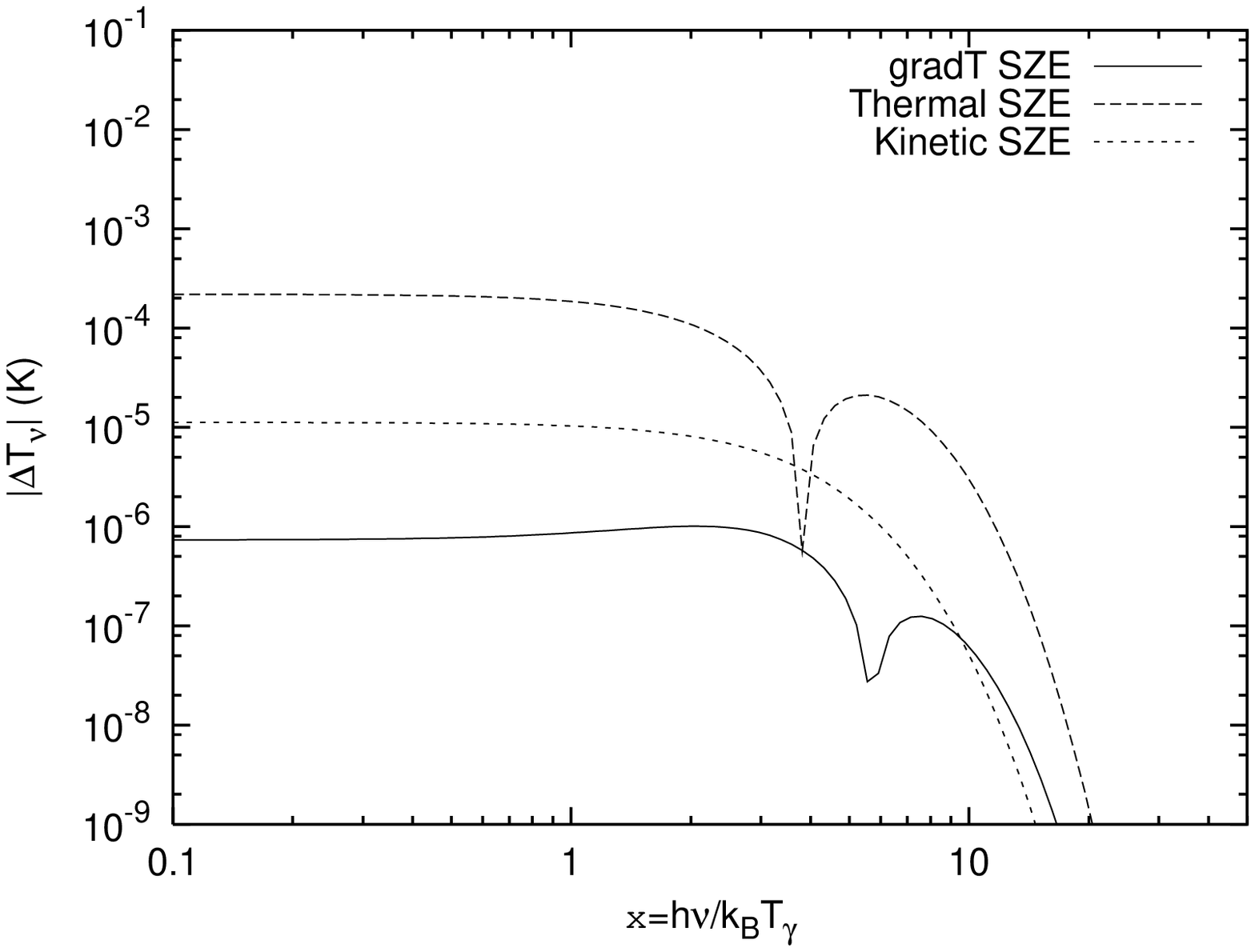}
\end{center}
\caption{Comparison of the CMB spectrum distortion due to the gradT SZE with the thermal
SZE and the kinetic SZE.  The absolute amplitude of 
each effects are compared by log-linear plot. 
A vertical axis is described in K. A  horizontal axis is the frequency of the CMB photons
normalized by its temperature. The solid line shows the spectrum of the gradT SZE with amplitude shown in Eq.(6). The dashed line shows
the spectrum of the thermal SZE assuming an uniform sphere of radius 0.5Mpc 
with electron density of 
$10^{-3}{\rm cm^{-3}}$ and 
electron temperature of 10keV. 
The dotted line shows the spectrum of the kinetic SZE assuming 
the same uniform sphere moving with a peculiar velocity
of 600${\rm km/s}$. }

\end{figure}

\begin{figure}[htbp]

\begin{center}
\includegraphics[width=0.7\linewidth]{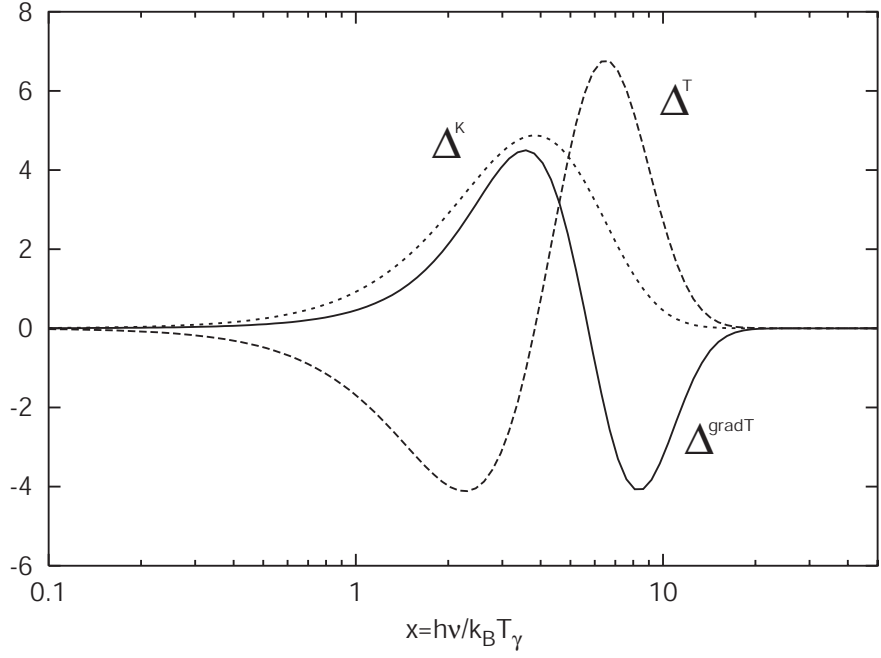}
\end{center}

\caption{The kernel functions of 
the three SZE spectrums are shown. 
The solid line shows the kernel function of the gradT SZE. 
Dashed line and dotted line are the kernel function of the thermal SZE, $\Delta^T(x)=x^4 e^x/(e^x-1)^2 
(x {\rm coth}(x/2)-4)$, and the kinetic
SZE, $\Delta^K(x)=x^4 e^x/(e^x-1)^2$, 
respectively\cite{Birk99}.  A vertical axis is described in dimensionless 
arbitrary unit.}
\end{figure}

\begin{figure}[htbp]
\begin{center}
\includegraphics[width=1.0\linewidth]{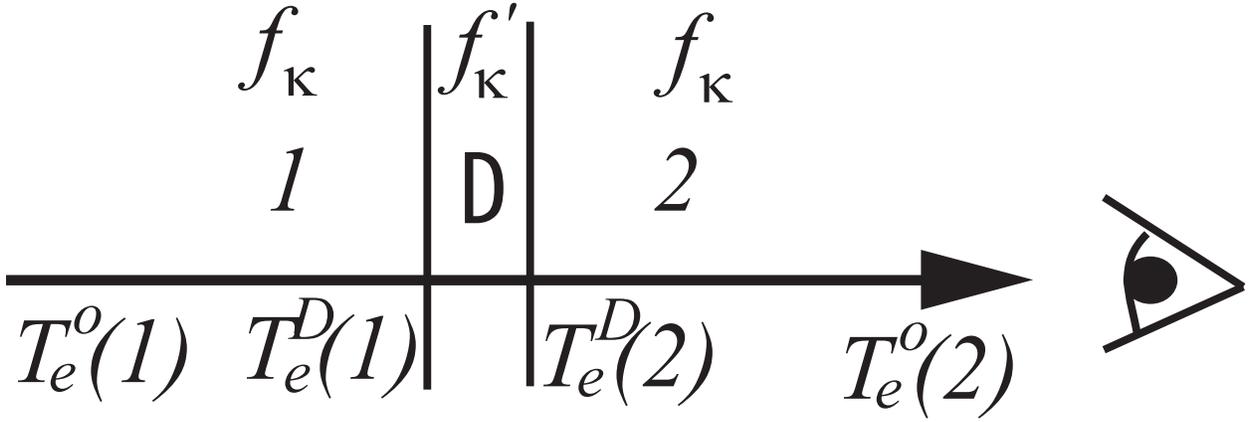}

\end{center}
\caption{An illustration of the line of sight which passes one discontinuity.
The line of sight toward an observer is illustrated by an long arrow. 
The region denoted by $D$ is a discontinuity. The regions denoted by 1 and 2 
are the regions where the temperature and density distributions of the hot gas 
are smooth. The electron temperatures at the edges of the discontinuity is 
denoted by the super script $D$ and the electron temperature at the outer 
edges of the cluster is denoted by the super script $o$, respectively.  The fractional heat conductivity in the smooth region is described by $f_{\kappa}$ and 
that in the discontinuity is described by $f'_{\kappa}$, respectively.}
\end{figure}

\end{document}